\documentclass[twocolumn,secnumarabic,amssymb, nobibnotes, aps, prd]{revtex4-2}

\newcommand{\macro}[1]{\texttt{\textbackslash#1}}
\newcommand{\m}[1]{\macro{#1}}

\usepackage{amsmath}
\usepackage{epsfig}
\usepackage{epic}
\usepackage[english]{babel}
\usepackage{color}
\usepackage{verbatim}
\definecolor{azul}{rgb}{0.1,0.2,0.6} % valores de las componentes roja, verde y azul (RGB)
\definecolor{verde}{rgb}{0.1,0.5,0.3}
\definecolor{bordo}{rgb}{0.9,0.3,0.3}
\usepackage{lipsum}
\usepackage{slashed}
\usepackage{tikz-cd}
\usetikzlibrary{automata}
\usetikzlibrary{topaths}
\usepackage{pstricks}
\usetikzlibrary{automata,positioning,calc}
\usetikzlibrary{arrows}

\usepackage{mathrsfs}
\usepackage[scr=dutchcal,calscaled=1]{mathalfa}
\usepackage{wasysym}

\usepackage{relsize} % bloques de tamaño variable
\usepackage{amsmath,amssymb}
\usepackage[hidelinks]{hyperref}

\begin{document}

\title{\Large \sc equivalent quantum systems}
\author{M. Caruso}%
\email{mcaruso@fidesol.org}
\affiliation{Fundación I+D del Software Libre, \href{http://www.fidesol.org}{$\mathtt{FIDESOL}$}, Granada (18100), España}
%\affiliation{%Departamento de F\'isica Te\'orica y del Cosmos, 
%Campus de Fuentenueva, Universidad de Granada, Granada (18071), Espa\~na.}
\affiliation{Universidad Internacional de La Rioja - $\mathtt{UNIR}$, Spain}
\email{mariano.caruso@unir.net}

\begin{abstract}{
We have studied quantum systems on finite-dimensional Hilbert spaces and found that all these
systems are connected through local transformations. Actually, we have shown that these transformations give rise to a gauge group that connects the hamiltonian operators associated with each
quantum system. This bridge allows us to connect different quantum systems, in such way that studying one of them allows to
understand the other through a gauge transformation. Furthermore, we included the case where the 
hamiltonian operator can be time-dependent. An application for this construction it will be achieved in the theory of control quantum systems.  %This result can be applied to the field of simulation of quantum systems, in order to  mimic more complicated quantum systems from another simulatable quantum system.
}
\end{abstract}
\vspace{0.3cm}

%\date{\today}

\maketitle

\begin{comment}
The bridge between $\pmb{\varphi}$ and $\pmb{\varphi}'$ is built through a local transformation $\pmb\omega$. Also the correspond link between $\pmb{H}$ and $\pmb{H}'$ through a local transformation $\pmb\omega$ is given by
\begin{equation}\label{b} 
\pmb{H}_+=\pmb{\omega}\; \pmb{H}_- \pmb{\omega}^{-1}+  i(\partial_t\pmb{\omega})\pmb{\omega}^{-1},
\end{equation}
for all $\nu$ and $\mu$. The expression \eqref{b} is almost equal to \eqref{a} but with an added term $i(\partial_t\pmb{\omega}) \pmb{\omega}^{-1}$.

We will see that the group of this kind of transformations is structured as a gauge group. 
We can formalize and generalize this idea in the following sections and we prove that is possible to connect any pair of hamiltonian operators $(\pmb{H},\pmb{H}')$ and any pair of complex vectors $(\pmb{\varphi},\pmb{\varphi}')$, associated with these, in a similar way that \eqref{b} and \eqref{demostrar}, respectively
\begin{align}
&\pmb{H}'=\pmb{\omega} \pmb{H} \pmb{\omega}^{-1}+ i \partial_t\pmb{\omega}\; \pmb{\omega}^{-1},\nonumber \\
&\\
&\pmb{\varphi}'=\pmb{\omega}\,\pmb{\varphi}.\nonumber   
\end{align}

\end{comment}

\section{Introduction}

In this work, we have developed a procedure to connect a given pair of quantum systems via a local transformation. We describe specifically a map among the respective Hilbert spaces that connect its vector objects (which represent quantum states) and its hamiltonian operators. We will studied the case in which the corresponding Hilbert spaces are finite-dimensional, but this results can be enunciated for infinite, but countable, dimensional Hilbert spaces. This correspondence is a useful tool to map quantum systems in order to study one of them through the other one.

%A productive and promising field to apply these ideas may be the \textit{quantum simulations}. 
At the end of 20th century, R. Feynman asked the following question: \textit{What kind of computer are we going to use to simulate physics?} [...] \textit{I present that as another interesting problem:
to work out the classes of different kinds of quantum mechanical systems which are really intersimulatable $-$which are equivalent$-$} [...] \textit{The same way we should try to find out what kinds of quantum mechanical systems are mutually intersimulatable, and try to find a specific class, or a character of that class which will
simulate everything} \cite{Feynman}.

We will prove that any two quantum systems on respective Hilbert spaces which are  finite dimensional are connected via a gauge transformation. This includes the case in which any of its corresponding hamiltonian may be time dependent.  We intend to open a way to establish the equivalence class previously mentioned by Feynman \cite{Feynman}. The rest of the article is organized as follows: we present a brief mathematical description of a general quantum system in a denumerable Hilbert space are described in section II. A motivation of the problem is described in section III. The formal aspects of the equivalence between quantum systems is shown in section IV. 
An application of these formal ideas in the control quantum systems area is exposed in section VI, is a particular approach in adiabatic  
A reduction algorithm for a sum of hamiltonian operators is presented in section VII and finally the conclusions and final comments are presented in section VIII.

\section{Quantum system on a denumerable hilbert space}

We will start by reviewing the basics aspects of quantum systems. Starting to finite numerable Hilbert space, let us consider a general quantum system Q which can be described in a certain $n-$dimensional Hilbert space $\mathscr{H}_n$. The \textit{deterministic} temporal evolution of a quantum system is driven by a hamiltonian operator $H$ (eventually time-dependent) defined on $\mathscr{H}_n$ \cite{Hall}. This operator modifies the vector state $|\psi(t)\rangle$  at time $t\in \mathcal{T}\subseteq\mathbb{R}$, by the equation
\begin{equation}\label{schr eqt}
i\partial_t |\psi(t)\rangle=H|\psi(t)\rangle,
\end{equation}
where $\partial_t$ represents the partial time derivative, leaving the possibility that such states may depend on other independent quantities. Note that a partial time derivative is used because $|\psi(t)\rangle$ can be dependent of other quantities. The equation \eqref{schr eqt} is written in natural units, e.g. $\hbar=1$.  Note that if the solution of \eqref{schr eqt} depends on time only or just other quantities that also depend on time, the partial derivative should be changed to total derivative $d_t$. In this paper the implementation of partial time derivative will be the same.

A solution of \eqref{schr eqt} is expressed as a $t-$parametrized curve on $\mathscr{H}_n$:
\begin{equation}\label{curveket}
|\psi\rangle : \mathcal{T}\longrightarrow \mathscr{H}_n.
\end{equation}
Using an orthonormal fixed basis \label{la base esta} of $n-$states: $\pmb{\beta}{=}\{|\psi_k\rangle\}_{k\in I_n}$, where $I_n=\{1,\cdots,n\}$ is the set of the first $n$ natural numbers, it is possible to represent the equation \eqref{schr eqt}  and its solution \eqref{curveket}. The inner product defined in the Hilbert space $\mathscr{H}_n$ allow us to express the state of the system at time $t$, $|\psi(t)\rangle$, in terms of its coordinates in the basis $\pmb{\beta}$ as 
\begin{equation}\label{coordinates}
\varphi_k(t):=\langle \psi_k |\psi(t)\rangle, \quad \forall k\in I_n.
\end{equation}
Note that the \textit{bra-ket} notation  is used to denote the inner product in $\mathscr{H}_n$, $\langle \star | \ast\rangle :\mathscr{H}_n\times \mathscr{H}_n \longrightarrow \mathbb{C}$. Thus, we have a time-parametrized curve on $\mathbb{C}^n$
\begin{equation}\label{curvecomplex}
\pmb{\varphi}:\mathcal{T}\longrightarrow \mathbb{C}^n,
\end{equation}
where $\pmb{\varphi}$ is written  in terms  of the coordinates of $|\psi\rangle$ in base $\pmb{\beta}$
\begin{equation}\label{vectorstate}
\pmb{\varphi}=(\varphi_1,\cdots,\varphi_n)^\mathfrak{t},
\end{equation}
where $\mathfrak{t}$ is the matrix transposition. Both curves \eqref{curveket} and \eqref{curvecomplex} refers to the same quantum system but different spaces $\mathscr{H}_n$ and $\mathbb{C}^n$, respectively. The  expression \eqref{coordinates} associates each element of the basis $\pmb{\beta}$ of $\mathscr{H}_n$  to one element of the canonical, or standard, basis of $\mathbb{C}^n$, i.e. a set of vectors $\{\pmb{e}_k\}_{k\in I_n}$ such that $\pmb{e}_k=(0,\cdots,1_{(k)},\cdots,0)$. In summary each $ |\psi_k\rangle$ corresponds only to $\pmb{e}_k$, for each $k\in I_n$.

Also, the complex vector curve \eqref{curvecomplex} satisfies another version of the equation \eqref{schr eqt}, given by
\begin{equation}\label{schr eqt v2}
i\partial_t \pmb{\varphi}(t)=\pmb{H\varphi}(t),
\end{equation}
where  $\pmb{H}{\in}\mathbb{C}^{n\times n}$ is a complex matrix that represents the hamiltonian operator $H$ in the basis 
$\pmb{\beta}$ and whose matrix elements are denoted by 
\begin{equation}\label{hamiltonain elements}
H_{kl}=\langle \psi_k |H| \psi_l\rangle
\end{equation}
In this manuscript, we refer to the hamiltonian operator, or hamiltonian matrix simply as hamiltonian.

\section{Promoting the general problem}
We are interested in finding a connection between
a given pair of quantum systems (Q,Q$'$) whose states belongs to their respective Hilbert spaces. Firstly we considered that they have the same dimension, namely $n$. In such case, we take two basis $(\pmb{\beta},\pmb{\beta}')$ associated with the pair of Hilbert spaces $(\mathscr{H}_n,\mathscr{H}'_n)$ that allows us to obtain two hamiltonian matrices $(\pmb{H},\pmb{H}')$ both in $\mathbb{C}^{n\times n}$. The dynamics of each quantum system is regulated by two equations similar to \eqref{schr eqt v2}. We denoted its solutions by $\pmb{\varphi}$ and $\pmb{\varphi}'$, both in $\mathbb{C}^n$, which are the respective wave functions associated to each Hilbert space basis $\pmb{\beta}$ and $\pmb{\beta}'$, defined by the expression \eqref{coordinates}, \eqref{curvecomplex} and \eqref{vectorstate}. In general terms, we expected that $\pmb{\varphi}$ and $\pmb{\varphi}'$ be related by
\begin{equation}\label{linearmapping}
\pmb{\varphi}'=\pmb{\omega\varphi}.
\end{equation}
in principle $\pmb{\omega}$ is an non$-$singular matrix in order to obtain the inverse connection. Also, this connection must be a linear in order to preserve the linear structure of equation \eqref{schr eqt v2}. On the other hand, we consider a connection between its hamiltonians $\pmb{H}$ and $\pmb{H}'$ through a certain map that depends on $\pmb{\omega}$, denoted by
\begin{equation}\label{connectionHH'}
\pmb{H}'=\pmb{\Omega}_{\pmb{\omega}}(\pmb{H}).
\end{equation}
For \textit{diagonalizable} hamiltonian matrices $(\pmb{H},\pmb{H}')$ which have the same spectrum, they are connected by a similarity transformation: $\pmb{H}'=\pmb{\omega}\pmb{H}\pmb{\omega}^{-1}$, but this case is very restrictive. In order to include a more general situation between any pair of quantum systems, we need to study other options for a general mapping $\pmb{\Omega}_{\pmb{\omega}}$ beyond the similarity transformation. We will see that it is even possible to connect quantum systems even in the case in which its hamiltonians do not share the same spectrum.  This article goes a lot further, exploring the idea of how different the hamiltonian operators can be connected, so that, if one of them is soluble we can use said solution to obtain the solution of the other via this mapping.

However, we want to clarify that this problem cannot be completely solved using the idea of quantum pictures, e.g. Schrödinger, Heisenberg and Interaction, formalised by Dirac \cite{Dirac}. We must to emphasize that in these cases: when going from one \textit{picture} to another we are dealing with the same quantum system, in fact the term picture perfectly reflects the idea of \textit{seeing} the same quantum system, but from another frame or perspective. Now, this article pretend to argue, that given two quantum systems $Q$ and $Q'$ (eventually different) with Hamiltonian operators $\pmb H$ and $\pmb{H}'$ there is a mapping this two quantum systems. This much more than a change of picture or representation of the same quantum system, since it necessarily implies the existence not only of the mentioned images but also the possibility of connecting very different quantum systems.

%%%%%%%%%%%%%%%%%%%%%%%%%%%%%%%%%%%%%%%%%%%%%%%%%%%%%%%%%%%%%%%%%%%%%%%%%%%%%%%%%%%%%%%%
\section{Formal aspects of equivalent quantum systems}\label{formal asp. equiv.}
%\textcolor{bordo}{\section{Formal aspects of these \\ gauge transformations}} 
%\textcolor{bordo}{ El t\'ermino gauge ¿debe quedar claro desde antes? \\O bien el siguiente t\'itulo}
%\textcolor{bordo}{\section*{Formal aspects of the map $\large\pmb{\Omega_\lambda}$}} 

We considered a map $\pmb{\Omega}_{\pmb{\omega}}$,  given a nonsingular matrix $\pmb{\omega}$, which transforms a matrix $\pmb{H}\in \mathbb{C}^{n\times n}$ as
\begin{equation}\label{map}
\pmb{\Omega}_{\pmb{\omega}}(\pmb{H}) = \pmb{\omega} \pmb{H} \pmb{\omega}^{-1} +i  (\partial_t\,\pmb{\omega}) \pmb{\omega}^{-1},
\end{equation}
where $\pmb{\omega}(t)$ is a $t-$differentiable non$-$singular matrix of $n\times n$, i.e.  $\pmb{\omega}:\mathcal{T}\longrightarrow\mathsf{GL}(n,\mathbb{C})$, also $\pmb{\Omega}_{\pmb{\omega}}\in \mathbb{C}^{n\times n}$.  The map $\pmb{\Omega_\omega}$ \eqref{map} is composed by a similarity transformation of $\pmb{H}$, defined by $\pmb{\omega}$, plus another time-dependent term. We will prove that the collection of this transformations $\{\pmb{\Omega}_{\pmb{\omega}}: \forall\, \pmb{\omega}\in \mathsf{GL}(n,\mathbb{C})\}$  form a group of local (gauge) transformations, with the composition of maps as a single associative binary operation. The locality of the transformation is due to the $t-$dependence of $\pmb{\omega}$. 

%There expression  \eqref{map} can be written in a more compact way\cite{lutkepohl}\begin{equation}\label{alt map}\pmb{\Omega}_{\pmb{\omega}}(\pmb{H}) = \pmb{\omega} \pmb{H} \pmb{\omega}^{-1} +i  \partial_t\ln (\pmb{\omega}).\end{equation}
This kind of mapping was studied in previous works from a pure mathematical point of view for applications to differential equations in complex variables with singular operators \cite{Varadarajan1,Varadarajan2,Varadarajan3}. For a physical point of view the same kind of mapping was presented in \cite{Mustafa1,Mustafa2,Mustafa3} in order to solve particular quantum systems. Respect to that, in this section we studied the possibility to connect any pair of hamiltonian $(H,H')$ operators defined on their respective $n-$dimensional Hilbert spaces $(\mathscr{H}_n,\mathscr{H}_n')$; this hamiltonians are represented by the matrices $(\pmb{H},\pmb{H}')$ eventually time dependent. We proved that there is a non singular matrix $\pmb{\omega}$, $t-$dependent and differentiable, that connect $\pmb{H}$ and $\pmb{H}'$  in this way
\begin{equation}\label{connect}
\pmb{H}'=\pmb{\Omega}_{\pmb{\omega}} (\pmb{H}).
\end{equation}

If we composed two transformations $\pmb{\Omega}_{\pmb{\omega}}\circ\,\pmb{\Omega}_{\pmb{\omega}'}$ with
$\pmb{\omega}$ and $\pmb{\omega}'$ are nonsingular, we see that 
\begin{align}\label{composegamma}
\pmb{\Omega}_{\pmb{\omega}}\circ\,\pmb{\Omega}_{\pmb{\omega}'}\!
& = \pmb{\Omega}_{\pmb{\omega.}\pmb{\omega}'}.
%\pmb{\Omega}_{\pmb{\omega}}\circ\,\pmb{\Omega}_{\pmb{\omega}'}(\pmb{H}) & = \pmb{\Omega}_{\pmb{\omega.}\pmb{\omega}'}(\pmb{H}).
\end{align}

From expression \eqref{composegamma} we see that if 
\begin{equation}\label{commutator}
[\,\pmb{\omega,\omega}']=\pmb{0}\Longrightarrow [\,\pmb{\Omega_{\omega}},\pmb{\Omega}_{\pmb{\omega}'}]
%\pmb{\Omega_{\omega}}\circ\,\pmb{\Omega}_{\pmb{\omega}'}(\pmb{H})=\pmb{\Omega}_{\pmb{\omega}'}\circ\,\pmb{\Omega}_{\pmb{\omega}}(\pmb{H}).
\end{equation}

Using the properties of composition \eqref{composegamma} and \eqref{commutator},  we present an expression for the inverse map  $\pmb{\Omega}_{\pmb{\omega}^{-1}}$. First of all, we have trivially
\begin{equation}
\pmb{\Omega_{I}}=\pmb{I},
\end{equation}
where $\pmb{I}$ is the identity matrix.  If we consider the composed transform $\pmb{\omega}''=\pmb{\omega.\,\omega}'$ such that $\pmb{\omega.\,\omega}'=\pmb{I}=\pmb{\omega}'\pmb{.\,\omega}$ then from \eqref{commutator} we have 
\begin{equation}\label{unique inv} 
\pmb{\Omega_{\omega}}\circ\,\pmb{\Omega}_{\pmb{\omega}'}=\pmb{I}=\pmb{\Omega}_{\pmb{\omega}'}\circ\,\pmb{\Omega}_{\pmb{\omega}}.
\end{equation}
From \eqref{unique inv} we obtain a \textit{unique} inverse of $\pmb{\Omega}_{\pmb{\omega}}$ given by 
\begin{align}\label{Gammainv}
\big[\pmb{\Omega_{\omega}}\big]^{{-}1}=\pmb{\Omega}_{\pmb{\omega}^{-1}}.
\end{align}
For more details of the properties of composition \eqref{composegamma} and inverse transformation \eqref{Gammainv} see Appendix. %\textbf{A1}. 

We demonstrated that for any pair of $n\times n$, eventually $t-$dependent and differentiable, matrices $\pmb{H}$ and $\pmb{H}'$  there exist a non-singular $n\times n$, $t-$dependent and differentiable matrix $\pmb{\omega}$ that connect them. For that we can define the following \textit{equivalence relation}:
\begin{align}\label{equivrel}
\pmb{H}'\sim \pmb{H}\Longleftrightarrow \exists \;\pmb{\omega}: \pmb{H}'=\pmb{\Omega}_{\pmb{\omega}}(\pmb{H}).
\end{align}
For more details about that expression \eqref{equivrel} is a well-defined \textit{equivalence relation} see Appendix. % \textbf{A1}.  
From the equivalence relation \eqref{equivrel} then  $\pmb{\omega}$ satisfies the differential equation:
\begin{align}\label{lambdaeq}
%i\,\pmb{\dot{\omega}}
i\partial_t\,\pmb{\omega}=\pmb{H}'\pmb{\omega}-\pmb{\omega H}.
\end{align}

% VERSION POSTA de la PAGINA 4

First of all, the solution of \eqref{lambdaeq} exists for the trivial cases $\pmb{H}=\pmb{0}$ and $\pmb{H}'=\pmb{0}$, i.e. we denoted by $\pmb{\omega}_1$ and $\pmb{\omega}_2$ the respective solutions for each case 
\begin{align}
i\partial_t\,\pmb{\omega}_1&=\pmb{H}'\pmb{\omega}_1,\label{l1}\\
i\partial_t\,\pmb{\omega}_2&=-\pmb{\omega}_2\pmb{H}.\label{l2}
\end{align}
We can obtain $(\pmb{\omega}_1,\pmb{\omega}_2)$ as iterative nonsingular solutions \cite{Magnus}. %\textcolor{bordo}{, for more details of this solution see Appendix \textbf{A2}} 
 The existence of solutions for equations \eqref{l1} and \eqref{l2} implies that $\pmb{\omega}_1$ and $\pmb{\omega}_2$ connect $\pmb{H}' \sim \pmb{0}$ and $\pmb{0} \sim \pmb{H}$, respectively. This implication is true from the definition of the equivalence relation. From the existence of solutions for \eqref{l1} and \eqref{l2} then we have 
\begin{align}
\exists\; \pmb{\omega}_1: \pmb{H}'=\pmb{\Omega}_{\pmb{\omega}_1}(\pmb{0})
&\Longleftrightarrow\pmb{H}'\!\sim \pmb{0},\label{transitive1}
\\
\exists\; \pmb{\omega}_2: \pmb{0}\,=\,\pmb{\Omega}_{\pmb{\omega}_2}(\pmb{H})
&\Longleftrightarrow\;\pmb{0}\sim\pmb{H},\label{transitive2}
\end{align} 
and from transitivity of the equivalence relation \eqref{equivrel} we have $\pmb{H}'\sim \pmb{H}$. This means that there is a given $\pmb{\omega}$ that $\pmb{H}'=\pmb{\Omega}_{\pmb{\omega}}(\pmb{H})$. 

We express the solution $\pmb{\omega}$ as a function of the solutions of \eqref{l1} and \eqref{l2}, $(\pmb{\omega}_1,\pmb{\omega}_2)$, respectively. We say that a solution $\pmb{\omega}$ built in this way is a \textit{transitive solution}, or \textit{composite solution}. This name will be clear in the construction procedure of the solution $\pmb{\omega}$. From \eqref{transitive1} and \eqref{transitive2} we see that the \textit{transitivity solution} is constructed from the composition of transformations $\pmb{H}'=\pmb{\Omega}_{\pmb{\omega}_1}(\pmb{0})$ and $\pmb{0}=\pmb{\Omega}_{\pmb{\omega}_2}(\pmb{H})$ as follows
\begin{align}\label{comp}
\pmb{H}'=\pmb{\Omega}_{\pmb{\omega}_1}\pmb{(}\pmb{\Omega}_{\pmb{\omega}_2}(\pmb{H})\pmb{)}
\end{align}
from the composition rule \eqref{composegamma} applied to \eqref{comp}
\begin{align}
\pmb{H}'=\pmb{\Omega}_{\pmb{\omega}_1\pmb{\omega}_2}(\pmb{H})
\end{align}
where the \textit{transitive solution} is given by
\begin{align}\label{lambdacomp}
\pmb{\omega}=\pmb{\omega}_1\pmb{\omega}_2.
\end{align}

We have demonstrated that for any pair of this kind of matrices $(\pmb{H},\pmb{H}')$, there is a nonsingular matrix $\pmb{\omega}$ that connects $\pmb{H}$ and $\pmb{H}'$ through the map $\pmb{\Omega_\omega}$, given by the expression \eqref{map}, this is
\begin{equation}\label{Qequiv}
\pmb{H}'=\pmb{\Omega_\omega}(\pmb{H}).
\end{equation}

Suppose now that this pair of matrices  $\pmb{H}$ and $\pmb{H}'$ are the hamiltonian operators of the following differential equations
\begin{align}\label{equations}
i\partial_t\pmb{\varphi}=\pmb{H}\,\pmb{\varphi},\quad 
i\partial_t\pmb{\varphi}'=\pmb{H}'\pmb{\varphi}',
\end{align}
finally, from \eqref{Qequiv} and \eqref{equations} we have 
\begin{align}\label{conlc}
\pmb{\varphi}'=\pmb{\omega}\pmb{\varphi}.
\end{align}
In summary, the connection between  $\pmb{H}$ and $\pmb{H}'$ can be found at the level of the solutions of \eqref{equations}, i.e. $\pmb{\omega}$ connects both hamiltonian matrices via $\pmb{H}'{=}\pmb{\Omega_\omega}(\pmb{H})$ \eqref{Qequiv} and also both solutions via $\pmb{\varphi}'=\pmb{\omega}\pmb{\varphi}$ \eqref{conlc}.

%If the \textit{starting point} is a self-adjoint operator $\pmb{H}$ which is transformed  via $\pmb{\Omega_\omega}$ in the \textit{direction} of an unitary $\pmb{\omega}$ then the operator $\pmb{H}'$, as a \textit{final point}, will also be self-adjoint operator given by \eqref{Qequiv}.

%%%%%%%%%%%%%%%%%%%%%%%%%%%%%%%%%%%%%%%%%%%%%%%%%%%%%%%%%%%%%%%%%%%%%

%\textcolor{azul}{\textbf{MUEVE ESPECTRO}\\
An important aspect of $\pmb{\Omega_\omega}$ mapping lies in the possibility of introducing an interaction for  the starting hamiltonian $\pmb{H}$. In particular, if  $\pmb{\omega}$ commutes with $\pmb{H}$, then from \eqref{map} we have $\pmb{H}'{ = } \, \pmb{H}+i(\partial_t\,\pmb{\omega})\, \pmb{\omega}^{-1}$, and the second term can be interpreted as an interaction operator $\pmb{H_i}:=i(\partial_t\,\pmb{\omega})\, \pmb{\omega}^{-1}$.  In this case, from $[\pmb{H},\pmb{\omega}]=0$ then $[\pmb{H},\pmb{\omega}^{-1}]=0$ and if also $\pmb{H}$ is time independent,  $\partial_t \pmb{H}=0$, then  $[\pmb{H},\partial_t\,\pmb{\omega}]=0$, and finally $[\pmb{H},\pmb{H}_{\pmb{i}}]=0$.

Note that, if the hamiltonian operators $(\pmb{H},\pmb{H}')$ are both hermitian, then $\pmb{\omega}$ is unitary and $\pmb{H_i}= i(\partial_t\,\pmb{\omega})\, \pmb{\omega}^{-1}$ is also hermitian. On the other hand, if  $(\pmb{H},\pmb{\omega})$ are hermitian and unitary, respectively, then $\pmb{H}'$ is also hermitian. We can compute exactly the evolution operator associated to $\pmb{H}'$ in a multiplicative factorization way \cite{Suzuki,Suzuki2}.

The role of $\pmb{\omega}$ can be the interpreted as a modification of the hamiltonian spectrum of $\pmb{H}$, even though it is degenerate, e.g. there are at least two different eigenvectors $\pmb{\varphi}_i$ and $\pmb{\varphi}_j$ are associated to the same eigenvalue, $\lambda_i{=}\lambda_j$, it is possible to choose $\pmb{\omega}$ such that the corresponding eigenvalues associated with the mentioned eigenvectors are not the same $\lambda_i'{\neq}\lambda_j'$. The interaction $\pmb{H_i}$ is the responsible for shifting the spectrum of the departure hamiltonian $\pmb{H}$.  This brings the possibility to control the separation width between two eigenvalues, e.g.  energy gap, in order to encode information and implement a qubit \cite{Nielsen,Benenti}.

In order to expose the presented ideas we contemplate the following non trivial example. In order to show explicitly what is the \textit{arrival} hamiltonian $\pmb{H}'$ connected  to the \textit{departure} hamiltonian $\pmb{H}$ through $\pmb{\omega}$ using \eqref{map} we considered a two-dimensional Hilbert spaces named $\mathscr{H}_2$, isomorphic to $\mathbb{C}^2$. Using the set of  $2\times 2$ matrices $\mathscr{B}{=}\{\sigma_\mu\}_{\mu=0,\,\cdots, \, 3}$ where $\sigma_0$ is the identity matrix and $\{\sigma_k\}_{k=1,2,3}$ are the Pauli matrices. The set $\mathscr{B}$ is a basis of $\mathbb{C}^{2\times 2}$, thus we can express $\pmb{H},\pmb{\omega}\,{\in} \mathbb{C}^{2\times 2}$ as a linear combination of the the elements of $\mathscr{B}$, where $\{(h_\mu,\omega_\mu)\}_{\mu=0,\,\cdots,\,3}$ are its coordinates respectively. From $[\,\pmb{\omega},\pmb{H}\,]\,{=}\,\pmb{0}$ then $\omega_k{=}\,\xi \,{\cdot} \,h_k$ for each $k{=}1,2,3$. Note that $\pmb{\omega}$ is in general time dependent and $\pmb{H}$ is time independent, thus $\xi$ must be a function of time and the commutative condition of $\pmb{\omega}$ and $\pmb{H}$  In order to obtain a compact expression is useful to call $\pmb{h}{=}(h_1,h_2,h_3)$ then the coordinates of $\pmb{H}'$, $\{h'_\mu\}_{\mu=0,\,\cdots,\,3}$, are given by
\begin{align}\label{h'}
\begin{aligned}
h'_0&=h_0+\tfrac{1}{2}i\, \partial_t \ln[det(\pmb{\omega})]\\
&\\
h'_k&=h_k\bigg[1+\frac{\xi\,\dot{\omega}_0-\dot{\xi}\,\omega_0}{i\,det(\pmb{\omega})}\bigg]
\end{aligned}
\end{align}
where $det(\pmb{\omega})=\omega_0^2-\xi^2\, \pmb{h\boldsymbol{\cdot}h}$ and $\pmb{h\boldsymbol{\cdot}h}=\sum_{i=1}^3h_i^2$. The expressions \eqref{h'} are the corresponding coordinates of $\pmb{H}'$ in basis $\mathscr{B}$, which are accessible from $\pmb{H}$ via $\pmb{\omega}$ as an additive perturbation $\pmb{H_i}=i (\partial_t\pmb{\omega})\, \pmb{\omega}^{-1}$.

\begin{comment}
The mapping \eqref{connect} in terms of the evolution operators $\pmb{U}$ and $\pmb{U}'$ is reduced  to
\begin{equation}\label{unitary}
\pmb{\mathscr{U}}'(t)= \pmb{\mathfrak{u}}(t)\,\pmb{\mathscr{U}}(t)
\end{equation}
where $\pmb{\mathscr{U}}(t-s):=\pmb{U}(t,s)$ and $\pmb{\mathscr{U}}'(t-s):=\pmb{U}'(t,s)$, because $\pmb{H}$ and $\pmb{H}'$ are time independent  \cite{Reed-Simon,DynamicSystem,MarsdenRatiuAbraham,AbrahamMarsden,ArnoldDiffEq} and $\pmb{\mathfrak{u}}(t-s):=\pmb{\omega}(t)\pmb{\omega}^\dagger(s)$. 
\end{comment}

%%%%%%%%%%%%%%%%%%%%%%%%%%%%%%%%%%%%%%%%%%%%%%%%%%%%%%%%%%%%%%%%%%%%%

\section{basis-independent scheme}\label{Gen. basis ind.}

There is a generalization of this kind of mapping between hamiltonian operators defined on a denumerable Hilbert space. %In this section we implement the mapping independently of the chosen basis.%, in order to apply this formalism to connect two hamiltonian operators given in different basis. 

In the previous section we have dealt with operators represented by matrices, in a base of elements of a certain Hilbert space. Once the aforementioned base was specified, there was no explicit record of it. This is another of the reasons why we will now do a treatment that does not require specifying a basis and at the time of doing so the nomenclature will be able to express it explicitly. We have used \textbf{bold} notation to denote matrices and vectors, now we will return to normal typography to refer to operator over the Hilbert space and its abstract vector elements are denoted using the \textit{bra-ket} Dirac's notation  \cite{Dirac,Hall}.

%%%%%%%%%%%%%

The ideas developed in Section \ref{formal asp. equiv.} in order to connect any two hamiltonians via a transformation \eqref{map} can be expressed now using the \textit{bra-ket} notation. Given two hamiltonian operators $(H,H')$, on two isomorphic Hilbert spaces $(\mathscr{H,H'})$, associated to a differential equation of the form \eqref{schr eqt}, it can be found a time dependent mapping given by the invertible operator $\omega\,{:}\,\mathscr{H}{\longrightarrow} \mathscr{H}'$ which connects the solutions of the respective Schrödinger equation of the form \eqref{schr eqt} and the aforementioned hamiltonian are related by 
\begin{align}\label{abstract mapping}
\begin{aligned}
|\psi'\rangle&=\omega|\psi\rangle\\ 
H'&=\Omega_\omega(H):=\omega \,H \,\omega^{-1}+i\,(\partial_t\omega)\, \omega^{-1}.
\end{aligned}
\end{align}
If the natural units are not used, will be able to define a similar map $\Omega_\omega$, by multiplying the second term on the right side of by $\hslash$. 

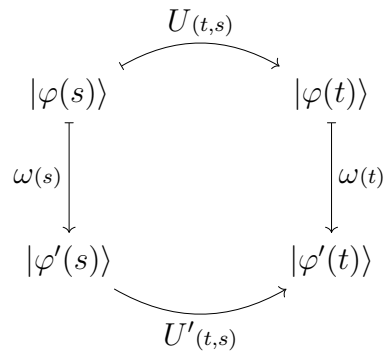
\begin{figure}[h!]
\[\large%\normalsize
\begin{tikzcd}
{|\varphi(s)\rangle} \arrow[rr, "{\mathlarger U(t,s)}", maps to, bend left] \arrow[dd, "{\mathlarger\omega(s)}"', maps to] &  & {|\varphi(t)\rangle} \arrow[dd, "{\mathlarger\omega(t)}", maps to] \\
&  &       \\
{|\varphi'(s)\rangle} \arrow[rr, "{\mathlarger{U'}(t,s)}"', bend right]                                              &  & {|\varphi'(t)\rangle}                                     
\end{tikzcd}\]
\caption{\footnotesize{This commutativity diagram shows how is the composition of transformations between equivalent quantum systems. The commutativity comes from the existence of the inverse of  $\omega(x)$ or the inverse of $U(y,z)$ for all $x,y,z$.}}\label{diagrama}
\end{figure}

In the general case these hamiltonians $(H,H')$ are not necessarily self-adjoint nor time-independent, the evolution operator of each quantum systems $(U,U')$ are related according to  $U'(t,s)=\omega(t)\,U(t,s)\, \omega^{-1}(s)$, for all $t,s\in \mathbb{R}$. In  Figure \ref{diagrama} a  commutative diagram shows how is the composition of this transformation. In case of $(H,H')$ are self-adjoint operators we have 
%we will call the first and second quantum systems of departure and arrival, respectively.
\begin{equation}\label{hibrid}
U'(t,s)= \omega(t)\,U(t,s)\,\omega^{\dagger}(s).
\end{equation}
where in this case $\omega$ is an unitary operator. 

A useful metaphor is to consider the launch of an \textit{abstract object} between two \textit{points}, corresponding to the two hamiltonians $H,H'$ and the throw is regulated  by $\omega$.  If $\omega$ is a unitary operator, then $\Omega_\omega$ is an endomorphism over the space of the self-adjoint operators (each of those are defined over the respective Hilbert spaces that have the same dimension $n$). Following the metaphor if the throw procedure is \textit{unitary} then the starting and finishing point will be \textit{self-adjoining}. This situation corresponds to a mapping two closed quantum systems. 

In this sense the operator $H$ will be called the \textit{departure} hamiltonian, wile the operator $H'$ will be considered as the target or \textit{arrival} hamiltonian. Respectively the corresponding quantum systems $Q$ and $Q'$ inherit such attributes and will be considered as the departure and arrival quantum systems. 

%%%%%%%%%%%%%

%In other words, when $\pmb{\omega}$ is an unitary matrix, then  $\pmb{\Omega_\omega}$ links up hermitian hamiltonian matrices. 
%in    this representation, and refer to \textit{closed} quantum systems. %, i.e. which are the systems that do not contain \textit{limbic} states
%A brief comment about the simulability of quantum systems in terms of their operator evolution is presented in \cite{Dirac}. 

% No bien justificado sobre la SIMULACION
%A general strategy to simulate the \textit{arrival} system Q$'$ consists the selection on the \textit{departure} system  Q and the connection between them, given by $\pmb{\omega}$. The departure system Q is chosen in order to simulate it analogically or digitally.  The bridge between Q and Q$'$, namely  $\pmb{\omega}$,  is not necessarily associated with any particular quantum system. For this reason the operator $\pmb{\omega}$ will be digitally simulated from a synthesis in quantum gates \cite{Giles Selinger}, which is an universal method. Therefore the expression \eqref{hibrid} could be implemented through a hybrid kind of simulation, that involves a digital-analog protocol for its implementation, as an indispensable requirement \cite{Solano}.

Until now we have considered the equivalence of quantum systems on Hilbert spaces with the same dimension or cardinality. We go one step further proving the equivalence of all quantum systems on a countable and finite-dimensional Hilbert space. Without loss of generality, we considered two Hilbert spaces $(\mathscr{H}_n, \mathscr{H}_{n'})$ where $n{<}n'$, and now the associated  hamiltonian  operators  $(H,H')$ have different dimension $(n,n')$, respectively. We can construct another hamiltonian $\widetilde{H}$ over $\mathscr{H}_{n'}$, associated with hamiltonian $H$ over the lower dimensional Hilbert space $\mathscr{H}_n$, defined by
\begin{align}\label{gen inf artif}
\widetilde{H}_{ij}=\Bigg{\{} \begin{matrix}
H_{ij}, &  & (i,j)\in I_n\times I_n\\
0, &  & (i,j) \notin I_n\times I_n.\\
\end{matrix}
\end{align}
\begin{comment}which in a given basis this elements \eqref{hamiltonain elements} can be represented in block form as
\begin{align}\label{gen inf artif2}
\pmb{\widetilde{H}}=\left(\begin{matrix}
\pmb{H}\,&    0    &   \cdots&0\\
0\,&  0 & \cdots&0\\
\vdots\,  & \vdots &\ddots&\vdots\\
0 \,& 0 &\cdots&0%
\end{matrix}\right)
\end{align}
\end{comment}
In such case, we know that there is a non-singular $\omega$,  now such that $H'{=}\,\Omega_\omega(\widetilde{H})$. This operator $\widetilde{H}$ corresponds to a new quantum system on a Hilbert space $\widetilde{\mathscr{H}}_{n'}$. We have completed the Hilbert space $\mathscr{H}_{n}$ with a number of $(n'-n)$ states, such that the resulting Hilbert space $\widetilde{\mathscr{H}}_{n'}$ and $\mathscr{H}_{n'}$ have the same cardinality $n'$. 

Now we present a physical interpretation of this procedure and some comments about the nature of this \textit{redundant} states. These additional states are collected in  a set, namely 
$\pmb{\alpha}=\big\{|\alpha_k\rangle{:}\, k\in I_{n'}-I_n\big\}$ 
%\tikz[baseline=(char.base)]{\node[shape=circle,draw,very thick,inner sep=0.9pt, black!70!](char){\footnotesize $\alpha_k$};} 
must be \textit{redundant} in the sense in which they are incorporated in order to form a Hilbert space $\widetilde{\mathscr{H}}_{n'}$, but they must not interact with the states of the Hilbert space $\mathscr{H}_n$ itself. This states do not modify the original dynamics on  $\mathscr{H}_n$. The states  of $\pmb{\alpha} $ (or $\pmb{\alpha}-$states) and the states of $\mathscr{H}_n$ are  \textit{mutually inaccessible}. If the quantum system was prepared in one of this $\pmb{\alpha}-$state, then the future state of this system cannot be left the initial state. Conversely, if the quantum system was prepared in a state that belongs to $\mathscr{H}_n$, then the future state of this system cannot be left $\mathscr{H}_n$ in order to go to $\pmb{\alpha}$. If we want that the dynamics on $\widetilde{\mathscr{H}}_{n'}$  corresponds to the dynamics on $\mathscr{H}_n$. In this way, the role  $\pmb{\alpha}-$states will be to complete the dimensionality of $\mathscr{H}_n$ and take it from $n$ to $n '$ in a dynamically \textit{innocuous} form. For all these reasons $\pmb{\alpha}-$states and the states of $\mathscr{H}_n$ are \textit{mutually inaccessible} and all $\pmb{\alpha}-$states are \textit{mutually inaccessible} as well.
 
\begin{figure}[h!]
\begin{center}
\centerline{\includegraphics[totalheight=3cm]{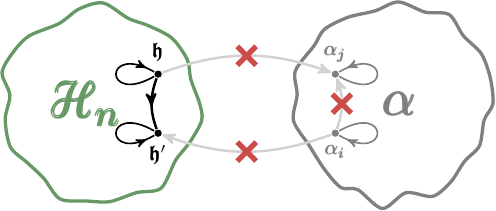}}
\end{center}
\caption{(Color online) \footnotesize{This diagram shows works the \textit{mutually inaccessible} relation for the $\pmb{\alpha}-$states and the states of $\mathscr{H}_n$. Those transitions with zero probability are indicated by the crosses (in \textcolor{bordo}{\textbf{red}})}.}\label{mut inacc}
\end{figure}

Let's see how these mentioned interpretations concerning the relations  between the $\pmb{\alpha}-$states and the states of $\mathscr{H}_n$ and the $\pmb{\alpha}-$states themselves, implies the hamitonian $\widetilde{H}$ from \eqref{gen inf artif}. For any $|\,\mathfrak{h}\,\rangle{\in}\, \mathscr{H}_n$ and  $|\alpha_i\rangle,|\alpha_j\rangle\,{\in}\,\pmb{\alpha}$, where $i{\neq}j$, the conditional probabilities % $P_{\alpha_i,\mathfrak{h}}$ and $P_{\mathfrak{h},\alpha_i}$,
 associated to the transitions $|\alpha_i\rangle{ \longmapsto} |\,\mathfrak{h}\,\rangle $ and $ |\,\mathfrak{h}\,\rangle{\longmapsto}|\alpha_i\rangle $,  are equal to zero. These two transition probabilities reveals the \textit{mutual inaccessibility} between $\pmb{\alpha}$ and $\mathscr{H}_n$. On the other hand, the conditional probabilities % $P_{\alpha_i,\alpha_j }$ 
associated with the transition $|\alpha_i\rangle{ \longmapsto} |\alpha_i\rangle $ are also equal to zero and reveals the \textit{mutual inaccessibility} of all $\pmb{\alpha}-$states themselves. These conditional probabilities come from the square modulus of complex \textit{matrix} elements of the evolution operator. Finally, given that the close relationship between this operator and the hamiltonian, implies the exact form of the hamiltonian  $ \widetilde{H}$ from \eqref{gen inf artif}. %Si hace falta se agrega esto % the conditional proability of Prob alpha--> H =0=Prob alpha--> H, estas primeras dos implican  and Palpha_j-->alpha i =delta ij. 
In  Fig. \ref{mut inacc} we have summarized the previous comments on the forbidden transitions between the states of $\mathscr{H}_n$ and $\pmb{\alpha}$.
%%%%%%%%%%%%%%%%%%%%%%%%%%%%%%%%%%%%%%%%%%%%%%%%%%%%%%%%%%%%%%%%%%%%%%%%%%%%%%%%%%%%%%%%%%%%%%%%%%%%%%%%%%%%%%%%%%
%%%%%%%%%%%%%%%%%%%%%%%%%%%%%%%%%%%%%%%%%%%%%%%%%%%%%%%%%%%%%%%%%%%%%%%%%%%%%%%%%%%%%%%%%%%%%%%%%%%%%%%%%%%%%%%%%%

\begin{figure}[h!]
\begin{center}
\centerline{\includegraphics[totalheight=6cm]{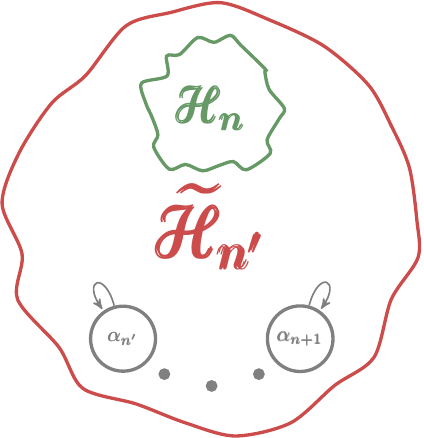}}
\end{center}
\caption{(Color online) \footnotesize{This diagram shows how is the composition of  $\widetilde{\mathscr{H}}_{n'}$ from $\mathscr{H}_n$ and a set of ($n'{-}n$) \textit{redundant} states.}}\label{complete}
\end{figure}

The linear combination of the elements of $\pmb{\alpha}$, or \textit{linear span}, contain the states that belong to $\widetilde{\mathscr{H}}_{n'}$ but does not belongs to $\mathscr{H}_n$, this is the complement of $\mathscr{H}_n$ in order to generate $\widetilde{\mathscr{H}}_{n'}$, namely $span(\pmb{\alpha})=\widetilde{\mathscr{H}}_{n'}\setminus\mathscr{H}_n$. For illustrative purposes Fig. \ref{complete} shows how is this composition.

%All the states of $\{$\tikz[baseline=(char.base)]{\node[shape=circle,draw,thick,inner sep=0.9pt, gray](char){\footnotesize $k$};} : $k\in[n+1,n']\subseteq \mathbb{N}\}$ are isolated or mutually disconnected and also from each state of $\mathcal{S}$. If the process start in some \tikz[baseline=(char.base)]{\node[shape=circle,draw,thick,inner sep=0.9pt, gray](char){\footnotesize $k$};} of this redundant set, it stays there forever. 

\section{CONTROL IN QUANTUM SYSTEMS}\label{quantum control}

A remarkable field to apply the presented tools is control theory of quantum systems, in order to implement a quantum simulator conceived as a controllable system whose aim is to mimic the static or dynamical properties of another quantum system \cite{Georgescu Review}. %\textcolor{verde}{Today this field can be roughly divided into digital quantum simulations, analog quantum simulations, and a combination of both. The digital quantum simulation, proposed by Lloyd \cite{Lloyd}, deals with the synthesis of a given operator evolution in \textit{quantum gates}. The advantage of this approach is the flexibility, which introduce the quantum error correction and, then, universality. One of its disadvantages is the large number of quantum gates that may be involved in the synthesis, which implies huge technological effort to maintain the coherence of the states. Whereas in the analog quantum simulations there is a \textit{first} quantum system that may be not experimentally easy realizable or controllable and a second quantum system that mimics the first one. The advantage of this approach is the possibility to describe quantum systems in larger Hilbert spaces. One of the main issues concerns to find a quantum system able to mimic certain aspects to simulate the first quantum system. And finally, there is a win-win strategy that consists of a hybrid digital-analog simulations to combine the best of both ideas \cite{Solano}. Focusing on the digital simulation of hamiltonian dynamics for quantum systems, we see that it is inevitable to deal with a numerable and even finite version of such quantum systems, i.e. the involved Hilbert spaces are finite-dimensional. On the other hand in the analog quantum simulations, we will focus on countable$-$dimensional Hilbert spaces. According to this, we must study quantum systems on a countable dimensional (denumerable) Hilbert spaces is essentially relevant. }
Emphasizing the controllability of a given quantum system, the mapping defined in \eqref{abstract mapping} can be useful to drive the evolution of this system. There are many approaches to control quantum systems, the formalism in this section it starts by choosing the desired state trajectory and then engineers a control that transport the system along this trajectory. 
 
Let's consider quantum system governs by a time dependent hamiltonian $H(t)$, rigorously speaking this is a time-indexed family of self-adjoint operators on the hilbert space $\mathscr{H}$. Suppose that  for each time $t$ its \textit{point spectrum} is the set $\{\lambda_n(t)\}_{n\in I}$ of eigenvalues, in this case all are different, where $I$ is a denumerable set such that $card(I){=}dim(\mathscr{H})$. Also there is an instantaneous basis of orthonormal eigenvectors of $H(t)$ with respect to the inner product $\langle \star| \ast\rangle:\mathscr{H}\times \mathscr{H} \longrightarrow \mathbb{C}$ \cite{Hall,Reed-Simon},  named $\pmb{\beta}_t{=}\{|\varphi_n(t)\rangle\}_{n\in I}$. The orthonormality of $\pmb{\beta}_t$ is expressed through the inner product $\langle\varphi_m|\varphi_n\rangle{=}\delta_{mn}$, for all $t{\in}\mathcal{T}$, where $\delta_{mn}$ is the \textit{Kroenecker delta} which is $1$ if $m{=}n$, and $0$ otherwise.

In general, there is a correspondence to each eigenvalue $\lambda_n(t)$ and its \textit{eigenspace} because by definition $\mathscr{H}_{n,t}{=}\Big{\{|}\varphi_t\rangle {\in} \mathscr{H}{:}\, H(t)|\varphi_t\rangle{=}\lambda_n(t)|\varphi_t\rangle\Big{\}}$, for $\lambda_m(t){\neq} \lambda_n(t)$ then $\mathscr{H}_{m,t} \cap \mathscr{H}_{n,t}=\{0\}$. The hole point spectrum is called non-degenerate, i.e. there are no linearly independent eigenvectors associated with the same eigenvalue. As all  eigenvalues are different, mathematically means that $dim(\mathscr{H}_{n,t}){=}1$ for each $n$ and $t$, and physically means if there is no  transition between elements of two proper spaces we will have the guarantee that the system will be dominated by a unique eigenvalue. Given that $\pmb{\beta}_t$ is a basis for each time $t$, therefore 
\begin{equation}\label{prop.ins.eigen.sol}
|\varphi(t)\rangle{=}\sum_{n\in I}\xi_n(t) |\varphi_n(t)\rangle
\end{equation}
is a proposed solution of \eqref{schr eqt} for this time dependent hamiltonian. The functions $\{\xi_n(t)\}_{n\in I}$ are obtained replacing the proposed solution in the equation  \eqref{schr eqt}
\begin{equation}\label{eq ham dep}
i\partial_t\xi_n{=}\lambda_n\xi_n-i\sum_{k\in I}\langle \varphi_n|\partial_t \varphi_k\rangle  \xi_k,
\end{equation} 
where the time dependence was omitted in order to simplify the notation. In particular, for an initial condition $|\varphi(0)\rangle{\in}\mathscr{H}_{m,0}$ then for $n{\neq} m$: $\xi_n(0){=}0$, but $\partial_t\xi_n(0){=-}\langle \varphi_n|\partial_t \varphi_m\rangle\big{|}_{t=0}$ which is non zero in general. For this reason, even if the system was initially belongs to the eigenspace $\mathscr{H}_{m,0}$, it cannot be guaranteed that a posteriori the state of the quantum system will remain in the same eigenspace, transitions will inevitably take place. If it is intended to control the system so that it remains in a certain eigenspace then let's see how to express $\langle \varphi_n|\partial_t \varphi_m\rangle$ in another way, since it quantifies the rate of the mentioned transitions. From $H|\varphi_m\rangle{=}\lambda_m|\varphi_m\rangle$, then $\partial_t H|\varphi_m\rangle{+}H|\partial_t\varphi_m\rangle{=}\partial_t\lambda_m|\varphi_m\rangle{+}\lambda_m|\partial_t\varphi_m\rangle$, closing on the left with the \textit{bra} $\langle \varphi_n|$ where $n{\neq} m$, finally % $\langle \varphi_n|\partial_t H|\varphi_m\rangle{+}\langle \varphi_n|H|\partial_t\varphi_m\rangle{=}\lambda_m\langle \varphi_n|\partial_t\varphi_m\rangle$
\begin{equation}\label{controlador trans}
\langle \varphi_n|\partial_t\varphi_m\rangle=\frac{\langle \varphi_n|\partial_t H|\varphi_m\rangle}{\lambda_m{-}\lambda_n}.
\end{equation} 
%\textcolor{bordo}{REVISAR where $*$ denotes the complex conjugate to cover the case where $H$ is not necessarily self-adjoint.}  
If $\langle \varphi_n|\partial_t \varphi_m\rangle{\simeq} 0$ for $n{\neq} m$ the approximate solution of \eqref{eq ham dep} for each $n{\in}I$ is given by $\xi_n(t){\simeq\,} {\exp}\Big{\{}i\int_0^t \pmb{\big{(}}{-}\lambda_n(s){+}i\langle \varphi_n(s)|\partial_s\varphi_n(s)\rangle\pmb{\big{)}}ds\Big{\}}\xi_n(0)$, 
replacing each coordinate $\xi_n(t)$ in \eqref{prop.ins.eigen.sol} an approximately solution of \eqref{schr eqt} is obtained. 
If $\langle \varphi_n|\partial_t \varphi_m\rangle$ is exactly equal to zero, it  corresponds to a non-interaction between states characterised by each states, represented by these instantaneous eigenvector. Only the state corresponds to $n{=}m$ evolves in a non trivial way and decoupled of the rest of states.  The quantity $\int_0^t i\langle \varphi_n(s)|\partial_s\varphi_n(s)\rangle ds$, which appears in addition to the familiar dynamical phase $-\int_0^t\lambda_n(s) ds$ governs the temporal evolution of the state of the system and  is also real, see the demonstration in Appendix A2, for this reason can be interpreted as another phase.  The relevance of this quantity is such that has its own name, called the \textit{Berry phase} \cite{BerryPhase84}. This kind of particular phases appears also in classical physics and, given that there is an underlying geometric structure, these are  called \textit{geometric phases} \cite{BerryPhase89}.

In order to inhibit the mentioned transitions looking the expression \eqref{controlador trans} the time variation of the hamiltonian is done slowly enough with respect to the difference of the corresponding eigenvalues. The slowly variation of the hamiltonian is related to a \textit{quasistatic} process or misnamed in the literature an \textit{adiabatic} process.  There is a connection between the rate of variation of $H$ in order to control the system state permanence in a particular eigenspace  $\mathscr{H}_{m,t}$ for a sufficient large time $t$, in this respect. In this section, let's remember that it is intended that the state trajectory is along the instantaneous eigenstates of the reference hamiltonian $H$. 

The above discussion shows that in the natural (i.e. uncontrolled) behaviour of the system it is not the possible guarantee non-transitions between eigenspaces of $H$  to do  unless the process is quasistatic, but this problem could be approached in another scenario or quantum basis. 

In this section we have been consider an orthonormal basis named \textit{dynamical} basis $\pmb{\beta}_t{=}\{|\varphi_n(t)\rangle\}_{n\in I}$ which contains the instantaneous eigenvectors of $H(t)$  and the other scenario will be built on the \textit{static} or fixed basis  $\pmb{\beta}{=}\{|\psi_n\rangle\}_{n\in I}$, to write down an unitary operator $\omega{=}\sum_{n\in I}\omega_n |\psi_n\rangle\langle\varphi_n|$, for simplicity the time dependence was omitted for each \textit{bra} $\langle\varphi_n(t)|$. Note that each term in the operator $\omega$ is another operator, $|\psi_n\rangle\langle\varphi_n|$, associates each \textit{ket} $|\psi\rangle$ to a new \textit{ket} $|\psi_n\rangle\langle\varphi_n|\psi\rangle$ we interpret as the vector multiplied by the scalar. Given that $\omega$ is unitary then $\omega_n{=}e^{if_n}$ where $f_n$ is a real quantity, in other words is the argument or the \textit{phase} of the complex number $\omega_n$. Including the case that $f_n$ is a function of time and mapping the hamiltoninan $H$ via $\Omega_\omega$ we get 
\begin{align}\label{omega(H)}
\begin{aligned}
\Omega_\omega(H)&=\omega\,H\,\omega^\dagger + i(\partial_t\omega)\,\omega^{\dagger}\\
&=\mathlarger{\sum}_{n\in I}\pmb{\big{(}}\lambda_n+i\langle\partial_t\varphi_n|\varphi_n\rangle-\partial_t f_n \pmb{\big{)}} \,|\psi_n\rangle \langle \psi_n|	\\
&	{+}\mathlarger{\sum}_{\substack{(m,n)\in I^2\\m\neq n}}{i}\, e^{i(f_m-f_n)} \langle\partial_t\varphi_m|\varphi_n\rangle \,|\psi_m\rangle \langle \psi_n|,
\end{aligned}
\end{align}
where $I^2:=I\times I$. Note that $\langle\partial_t\varphi_m |\varphi_n\rangle{=}-\langle\varphi_m |\partial_t\varphi_n\rangle$ thus the first sum in \eqref{omega(H)} can be drive the state of the system in order to satisfied the desired control goal, if $f_n{\equiv}0$ for each $n{\in}I$.  Nevertheless the second sum in \eqref{omega(H)} leads to transitions between states of the system. Any attempt to suppress these transitions requires taking the approximation $\langle\varphi_m |\partial_t\varphi_n\rangle\simeq 0$ for all $m\neq n$ and, given \eqref{controlador trans}, this is equivalent to taking the quasistatic approximation. 

Based on the decomposition of the $\Omega_\omega$ mapping for a sum of two operators $A$,  $B$ and a generic $\omega$ all over $\mathscr{H}$ 
\begin{equation}\label{idea primigenia}
\Omega_\omega(A{+}B)=\Omega_\omega(A)+\omega\,B\,\omega^{-1}.
\end{equation}
Its demonstration follows directly from the definition of the map $\Omega_\omega$ \eqref{abstract mapping}. Is possible to find another hamiltonian named $\widetilde{H}$ such that $\Omega_\omega(H+\widetilde{H})$ can be drive the state of the original system whose hamiltonian is $H$
\begin{equation}\label{shortcut Omega}
\Omega_\omega(H{+}\widetilde{H})\,{=}\mathlarger{\sum}_{n\in I}\pmb{\big{(}}\lambda_n+i\langle\partial_t\varphi_n|\varphi_n\rangle \pmb{\big{)}} \,|\psi_n\rangle \langle \psi_n|.	
\end{equation}
The details of this procedure are in the Appendix A2.

The idea that we have been pursuing is looking for a hamiltonian $\widetilde{H}$ such that added to our original hamiltonian $ H $ which provides an evolution like would be achieved if an adiabatic process were valid for $H$ via $\Omega_\omega(H{+}\widetilde{H})$.

In addition to those readers invaded by the anxiety that comes from waiting to finish reading this article to arrive at the aforementioned Appendix A2, here is another shortcut to the closed expression for the additive hamiltonian $\widetilde{H}$ 
\begin{align}\begin{aligned}\label{additive hamiltonian}
\widetilde{H}=
\,\mathlarger{\sum}_{n\in I}\;&\pmb{\big{(}}\partial_tf_n +i\langle\partial_t\varphi_n|\varphi_n\rangle \pmb{\big{)}} \,{|\varphi_n\rangle} {\langle \varphi_n|}+\\
\,\mathlarger{\sum}_{n\in I}\;&i|\partial_t\varphi_n\rangle \langle\, \varphi_n\,|,
\end{aligned}
\end{align}
the first sum corresponds to a diagonal operator and the second sum corresponds to a non-diagonal operator.

Note that the phases $\{f_n\}_{n\in I}$ in the operator $\omega$ for our exposition are identically zero, but this is not the case in other contexts. In \cite{BerryTrans.Driving} the operator $\omega$ is defined considering a particular static basis defined from the dynamical basis evaluated in $t=0$, i.e. $\pmb{\beta}{:=}\pmb{\beta}_0$, and the phases are $f_n(t){=}\int_0^t \pmb{\big{(}}{-}\lambda_n(s) {+} i\langle \varphi_n(s)|\partial_s\varphi_n(s)\rangle\pmb{\big{)}} ds$. 

There is another equivalent expression for $\widetilde{H}$  
\begin{align}\begin{aligned}\label{additive hamiltonian 2}
\widetilde{H}=\mathlarger{\sum}_{n\in I}  \;\big{(} \partial_t 	f_n \big{)} \,{|\varphi_n\rangle} {\langle \varphi_n|}& + \\
\;\mathlarger{\sum}_{\substack{(m,n)\in I^2\\m\neq n}}i
\frac{\langle \varphi_m|\partial_t H|\varphi_n\rangle}{\lambda_n{-}\lambda_m} &{|\varphi_m\rangle\langle \varphi_n|},
\end{aligned}
\end{align}
again, the first sum corresponds to a diagonal operator and the second sum corresponds to a non-diagonal operator.

%$|\varphi(t)\rangle{\simeq} {\exp}\Big{\{}{-i}\int_0^t \Big{[}\lambda_n(s){-}i\langle \varphi_k(s)|\partial_s\varphi_n(s)\rangle\Big{]}\Big{\}}\xi_n(0)|\varphi_n(t)\rangle$ 
%%% CAMBIO de SECCIÓN? 
%$i\dot{\xi}_k{=}\big{[}\lambda_k-i\langle\varphi_k|\partial_t \varphi_k\rangle \big{]}  \xi_k-i\sum_{n\in I-\{k\} }\langle \varphi_k|\partial_t \varphi_n\rangle  \xi_n$ 

The methodology which was described is called an \textit{adiabatic shortcut} and in particular we have argued the \textit{transitionless driving} protocol, for a clearly mathematical explanation see \cite{BerryTrans.Driving} and for an  equivalent approach, called \textit{counter-diabatic} see \cite{Demirplak.Rice}.
While in  \cite{Lewis} is possible to find and use a time invariant operator to solve \eqref{schr eqt} via a similar reverse engineering protocol. Remark that transitionless driving is one of the adiabatic shortcut protocols \cite{BerryTrans.Driving}, but there are many others summarized in  \cite{Shortcuts.Rev,Shortcuts.Chap}.

On the other hand, the transformation under $\Omega_\omega$ of $H$ multiplied by a number $u\in\mathbb{C}$ is 
\begin{equation}\label{mult u}
\Omega_\omega(u\,H)=u\,\Omega_\omega(H)-i(u-1)(\partial_t \omega)\omega^{-1}.
\end{equation}
The transformation  under $\Omega_\omega$ of a finite sum of hamiltonian operators is given
\begin{equation*}
\Omega_\omega\pmb{\Big{(}}\sum_{k=1}^NH_k\pmb{\Big{)}}=\sum_{k=1}^N \Omega_\omega(H_k)-i(N-1)(\partial_t\omega)\omega^{-1}.
\end{equation*}

Another result of $\Omega_\omega$ is the invariance under a particular linear combination. If the \textit{departure} hamiltonian $H$ is given by a convergent linear combination of hamiltonians $\{H_k\}_{k\in J}$, where $J$ is a denumerable index set, $\sum_{k\in  J}u_k{=}\,u$, when $u$ can be time dependent, then 
\begin{equation*}\label{suma ponderada}
\Omega_\omega\pmb{\Big{(}}\sum_{k\in J}u_k\,H_k\pmb{\Big{)}}=\sum_{k\in J}u_k\,\Omega_\omega(H_k)-i(u-1)(\partial_t\omega)\omega^{-1},
\end{equation*}
in particular, if $u\neq 0$ the parameters $\{w_k \}_{k\in J}$ can be rewritten as $u_k=u \, w_k$ and conclude that
\begin{equation}\label{convexidades conservadas}
\Omega_\omega\pmb{\Big{(}}\sum_{k\in J}w_k\,H_k\pmb{\Big{)}}=\sum_{k\in J}w_k\,\Omega_\omega(H_k),
\end{equation}
because $\{w_k \}_{k\in J}$ satisfied $\sum_{k\in J}w_k{=}1$. In other words, can be considered as \textit{weights}. The expression \eqref{convexidades conservadas} is preserved even if the weights are \textit{signed} $w_k{\geq}0$ for all $k{\in}J$, in this case is called a convex combination. This quantities called weights $\{w_k\}_{k\in J}$ are eventually responsible to control the time spend to simulate each part of the convex sum \cite{Reed-Simon,DynamicSystem}. In this context, the invariance under convex combination allows to transform each hamiltonian control problem $\{w_k,H_k\}_{k \in J}$ into another $\{w_k,\Omega_\omega(H_k)\}_{k \in J}$ but preserving the weights. 

%Also note that \eqref{convexidades conservadas} can be obtained from \eqref{mult u} and ...

\section{Swallowing Algorithm}

Inspired in the algorithm presented in \cite{BerrySequentially} conformed by a sequence of unitary transformation, there is a possibility to reduce a sum of $N$ hamiltonian operators from one hamiltonian operator by a finite sequence of operators $
\{\omega_k\}_{k=1,\,\cdots,N}$ non necessarily unitary and its  implementation of the corresponding sequence $\{\Omega_{\omega_k}\}_{k=1,\,\cdots,N}$.

Once more, based on the decomposition of the $\Omega_\omega$ mapping exposed in \eqref{idea primigenia}, if also demand that $\Omega_\omega(A){=}0$ we can reduce \eqref{idea primigenia} to $\Omega_\omega(A+B){=}\omega\,B\,\omega^{-1}$. Suppose we are interested in studying the temporal evolution of a quantum systems whose hamiltonian is  $\sum_{j = 1}^N H_j$ using the following procedure, we can \textit{swallow} each term of this sum sequentially. As these are bounded Hamiltonian operators, then it will be lawful to demand  for an invertible operator $\omega_1$ that $\Omega_{\omega_1}(H_1){=}0$, what is achieved if $\omega_1$ satisfies $i\, \dot{\omega}_1{=}-\omega_1\,H_1$. Therefore ${\Omega_{\omega_1}}\pmb{\Big{(}}\sum_{j=1}^N H_j\pmb{\Big{)}}{=}\sum_{j=2}^N H^{(1)}_j$, with $H^{(1)}_j{:=}\omega_1\,H_j\,\omega_1^{-1}$. It can be seen that $\Omega_{\omega_1}$ takes a sum of $N$  hamiltonians and returns a sum of $N-1$ hamiltonians which none of these are exactly any of those original sum but are equivalent via a similarity transformation given by $\omega_1$. This procedure can be repeated $ N-1 $ times until leaving a single hamiltonian or taking one more step the $N$-step and concluding the swallowing task leaving an identically null hamiltonian.

Firstly we define the similarity transformations of the hamiltonians for each $j,k\in\{1,\cdots,N\}$
\begin{align}\label{HamiltonianosSimilares}
\begin{aligned}
H_j^{(k)}&{:=}\,\omega_k\cdots \omega_1 H_j(\omega_k\cdots \omega_1)^{-1},\\
H_j^{(0)}&{:=}H_j,
\end{aligned}
\end{align}
where the second expression denotes the \textit{starting point}, $H_j^{(0)}$, of each hamiltonian $H_j$ to initialize the algorithm.  We have used a compact notation $\omega_k\cdots \omega_1$ to denote the ordered composition of this operators. 

Secondly, we suppose a sequence of invertible operators $\{\omega_k: k=1,{\cdots},N\}$ such that 
\begin{equation}\label{OmegasquedeglutenH's}
\Omega_{\omega_k}\Big{(}H_k^{(k-1)}\Big{)}=0.
\end{equation}

Finally, before giving an account of the \textit{swallowing} algorithm, let us note the following recurrence relationship that will be useful to construct it
\begin{equation}\label{recurrenciaintermedia}
%H_j^{(k+1)}=\omega_{k+1}H_j^{(k)}\omega_{k+1}^{-1}
H_j^{(k)}=\omega_{k}\,H_j^{(k-1)}\omega_{k}^{-1}
\end{equation}
 %y  $\pmb{\omega}_0{=}\pmb{I}$ 
Then we can think of a protocol of $N{-}1$ steps (or $N$ steps if we want total swallowing) so that, using \eqref{idea primigenia}, \eqref{HamiltonianosSimilares}, \eqref{OmegasquedeglutenH's} and \eqref{recurrenciaintermedia} we can say that $k$-step is given by
\begin{equation}\label{k-esimo paso}
\Omega_{\omega_k\cdots\, \omega_1}\Big{(}\sum\nolimits_{j=1}^N H_j\Big{)}=\sum\nolimits_{j=k+1}^N H^{(k)}_j
\end{equation}
for each $k{=}1,{\cdots},N$.
The left member is designed to swallow $k$ addends, so for the original $N$ addends will remain $N{-}k$.

Let's see the proof by finite induction on $k$. The first step $k=1$ is true from  \eqref{idea primigenia} with $A{=}H_1$ y $B{=}\sum_{j=2}^N H_j$, \eqref{HamiltonianosSimilares} and \eqref{OmegasquedeglutenH's}. Taking the $k$-step in \eqref{k-esimo paso} and apply $\Omega_{\omega_{k+1}}$ then 
\begin{align}\label{k+1-esimo paso}
\begin{aligned}
\Omega_{\omega_{k+1}}\bigg{(}\Omega_{\omega_{k}\cdots\, \omega_1}\Big{(}\sum_{j=1}^N H_j\Big{)}\bigg{)}&=\Omega_{\omega_{k+1}}\bigg{(}\sum_{j=k+1}^N H^{(k)}_j\bigg{)}\\
\Omega_{\omega_{k+1}\omega_{k}\cdots\, \omega_1}\Big{(}\sum_{j=1}^N H_j\Big{)}&=\Omega_{\omega_{k+1}}\Big{(}H^{(k)}_{k+1}\Big{)}+\\
&+\sum_{j=k+2}^N \omega_{k+1}H^{(k)}_j\omega_{k+1}^{-1}\\
\Omega_{\omega_{k+1}\cdots\, \omega_1}\Big{(}\sum_{j=1}^N H_j\Big{)}&=\sum_{j=k+2}^N H^{(k+1)}_j
\end{aligned}
\end{align}
which is exactly the statement which appear in \eqref{k-esimo paso} but for the $(k{+}1)$-step.

Inasmuch as the map $\Omega_\omega$ is invertible, then the presented protocol connection is in both ways: from a sum of $N$-hamiltonians and a unique hamiltonian and vice versa. We have omitted the existence of this reverse procedure from the title of the section, since it is a digestive process that does not evoke good mental images or pleasant sensations for the reader of this work.

\section{Final observations} \label{sec-5}

The aim of present work it was to prove that there is a way to modify the behavior of a known quantum system, in order to get information of another quantum system that, at least, has a difficulty to be resolved directly.

Even when the state space of each Hilbert space has different cardinality, it is still possible to establish a link via a local transformation. This connection could be used including an eventually $t-$dependence of any of these hamiltonian operators. 

In summary, we have shown how for a given pair of quantum systems, finite-dimensional Hilbert spaces and its respective hamiltonian: $(\mathscr{H},H)$ and $(\mathscr{H}',H')$ they could be linked via gauge (local) transformations $\omega$, that allow us to obtain $H'$ from $H$, via $\Omega_\omega$.

In addition, this method allows us to address a new problem from another known one, using a non-local modulation ($\omega$) of the well-known solution $|\varphi\rangle$, following  \eqref{abstract mapping}. We have not only shown that this is feasible to do through formal and constructive proof of the existence of that $\omega$. But also we have indicated what is the right way to do it: should be across a linear and local (i.e. time-dependent) operation. 
%\textcolor{bordo}{For a given quantum system Q we take the set of all the quantum systems that are related to this via $\sim$.}

%We intend to deal with the topic of quantum simulation from an alternative perspective,  starting from a more well-known one, in the sense that the last one can be analytically soluble and/or simulatable.

Respect to the simulation of a quantum system Q$'$ we search for some other system that imitates the behavior of Q as well as possible. In other words, we must perform a \textit{casting call} of quantum systems or \textit{actors} which can be very limited, because it is a hard task to find another Q one to simulate Q$'$. We wanted to use this equivalence between quantum systems to simulate \textit{another} quantum system connected with  Q$'$. But when we said \textit{another}, we want to say \textit{any other} quantum system which is connected with Q$'$ through $\omega$. The map $\Omega_\omega$ applied to a given hamiltonian $H$  in \eqref{map}  works as \textit{makeup} that allows any \textit{actor} Q, to  simulate the first quantum system Q$'$, a priori, if Q is connected with Q$'$ through $\omega$.  Following the metaphor, the equivalence between this quantum systems expands that catalogue of actors that can make a good performance in order to mimic another quantum system and becoming that casting call, a priori more efficient. 

Given that the equivalence relation $\sim$ defines an equivalent class, see Appendix, % \textbf{A1}
the set of all transformations forms a \textit{guild of actors} capable of simulating, a priori, any quantum system of such systems conglomerate, which can be fully explored by such set of transformations. In particular, the subset of mapping wich $\omega$ unitary forms a \textit{conservative guild of actors} capable of simulating quantum systems with selfadjoint hamiltonian.

Regarding to control theory of quantum systems we have shown that there is a strong connection with our mapping $\Omega_\omega$  since it can be implemented to design temporal evolutions required a priori from a primal hamiltonian $H(t)$. We have dealt with the case of an orthonormal basis of eigenvectors of $ H(t) $ whose point spectrum is non-degenerate transferring this problem to a static basis also orthonormal. From the concrete $\omega$ can be seen that both bases, dynamical and statical are bi-orthonormal, i.e. $ \langle \psi_n | \varphi_m\rangle{=} \delta_{nm} $. This property allows to guarantee the original identification between each vector of the dynamical basis with the corresponding eigenvalue also for the statical basis. This can be generalized introducing a more general $\omega$ operators.

A final comment in this regard could be the implementation of the formal equivalence between quantum and classical systems proposed in \cite{caruso} in order to simulate quantum systems through specific circuits. The advantage of using such classical systems is that their controllability is simpler than for quantum systems in general, in order to adequately guide its temporal evolution. This implementation open the possibility to expands this catalogue  of actors capable of simulating the quantum system $Q$  even more with classical actors, who usually do not play that role.

Further applications of this methodological connection could be applied to perform computer simulation of quantum systems in a new way.

%\vspace*{0.5cm}

\section*{Acknowledgments}

We thank to University of Granada and \href{http://www.fidesol.org}{\texttt{FIDESOL}} for the support and recall also the anonymous readers for their constructive criticism to this work.

\section*{Competing interest}
The author declares that there are no competing interests.
\vspace*{0.4cm}
\section*{Appendix}
\appendix*
\vspace*{0.5cm}
%\appendix
\setcounter{equation}{0}

%%%%%%%%%%%%%%%%%%%%%%%%%%%%%%%%%%%%%%%%%%%%%%%%%%%%%%%%%%%%%%%%%%%%%%%%%%%%%%%%%%%%%%%%%%%%%%%%%%%%%%%%

%%%%%%%%%%%%%%%%%%%%%%%%%%%%%%%%%%%%%%%%%%%%%%%%%%%%%%%%%%%%%%%%%%%%%%%%%%%%%%%%%%%%%%%%%%%%%%%%%%%%%%%%
\section*{\sc \pmb{A1. Some structural properties of map  $\pmb{\Omega_\omega}$}}

$\bullet$ {\sc  \footnotesize MAPPING COMPOSITION}\\

We composed two transformations $\Omega_{\omega}\circ\,\Omega_{\omega}'$ with
$\omega$ and $\omega'$ are nonsingular and then prove that 
\begin{equation}\label{composite}
\Omega_{\omega}\circ\Omega_{\omega'}(H)=\Omega_{\omega\omega'}(H).
\end{equation}
for all $H$. We calculate directly $\Omega_{\omega}\circ\,\Omega_{\omega'}(H)$, denoting it by $\pmb{\square}=\Omega_{\omega}\circ\,\Omega_{\omega'}(H)$ 
\begin{align*}
\pmb{\square}& {=} \Omega_{\omega}\pmb{(}\omega' H \omega'^{-1} {+}i(\partial_t\omega')\omega'^{-1}\pmb{)}\nonumber\\
& {=} \omega\big[\omega' H \omega'^{-1} {+}i(\partial_t\omega')\,\omega'^{-1}\big]\omega^{-1}
{+}i(\partial_t\omega)\,\omega^{-1}\nonumber\\
%& = \pmb{\omega}\pmb{\omega}' \pmb{H} (\pmb{\omega\,\omega}')^{-1}+i\pmb{\omega}(\partial_t\pmb{\omega}')\pmb{\omega}'^{-1}\pmb{\omega}^{-1}+i(\partial_t\pmb{\omega})\,(\pmb{\omega}'\pmb{\omega}'^{-1})\pmb{\omega}^{-1}\nonumber\\
& {=} \omega\,\omega' H (\omega\,\omega')^{-1}
{+}i\omega(\partial_t\omega')(\omega\,\omega')^{-1}{+}i(\partial_t\omega)\,(\omega'\omega'^{-1})\,\omega^{-1}
\nonumber\\
& {=} \omega\,\omega' H (\omega\,\omega')^{-1}
{+}i\omega(\partial_t\omega')(\omega\,\omega')^{-1}{+}i(\partial_t\omega)\,\omega'(\omega\,\omega')^{-1}\nonumber\\
& {=}\omega\omega' H (\omega\,\omega')^{-1}
{+}i\partial_t(\omega\,\omega')(\omega\,\omega')^{-1}\nonumber\\
\pmb{\square} & {=} \Omega_{\omega\omega'}(H).\nonumber
\end{align*}
This completes the demonstration of \eqref{composite}: $\Omega_{\omega}{\circ}\;\Omega_{\omega'}{=}\;\Omega_{\omega\omega'}$ and expression \eqref{composegamma} is satisfied.\\

$\bullet$ {\sc \footnotesize INVERSE MAPPING}\\

We calculate explicitly $\Omega_{\omega^{-1}}(H)$ and then prove that 
\begin{equation}\label{invers}
\Omega_{\omega^{-1}}(H)=\Omega_{\omega}^{-1}(H),
\end{equation} 
for all $H$. Let's calculate the left hand of \eqref{invers}
\begin{align*}
\Omega_{\omega^{-1}}(H)
&=\omega^{-1} H \omega +i(\partial_t\omega^{-1})\omega\nonumber\\
&=\omega^{-1} H \omega +i(\partial_t\omega^{-1})\omega\nonumber\\
\Omega_{\omega^{-1}}(H)
&=\omega^{-1} H \omega -i\omega^{-1}\partial_t\omega.
\end{align*}

Finally we check directly that $\Omega_{\omega^{-1}}$ is equal to $\Omega_{\omega}^{-1}$, if for all $H$ we take $\pmb{\hexagon}=\Omega_{\omega}\circ \Omega_{\omega^{-1}}(H)$ and calculate
\begin{align*}
\pmb{\hexagon}&=\Omega_{\omega}(\omega^{-1} H \omega -i\omega^{-1}\partial_t\,\omega)\nonumber\\
&=\omega (\omega^{-1} H \omega -i\omega^{-1}\partial_t\omega)\,\omega^{-1}+ i (\partial_t\omega) \,\omega^{-1}\nonumber\\
&=\omega\, \omega^{-1} H \omega\, \omega^{-1} -i\omega\,\omega^{-1}(\partial_t\omega)\,\omega^{-1}+ i (\partial_t\omega)\, \omega^{-1}\nonumber\\
&= H-i(\partial_t\omega)\,\omega^{-1}+i  (\partial_t\omega)\, \omega^{-1}\nonumber \\
\pmb{\hexagon}&= H,
\end{align*}
where  $\partial_t(\omega^{-1}\omega){=}\,\pmb{0}$ then $\partial_t(\omega^{-1})\,\omega=-\omega^{-1}\partial_t\,\omega$. 
This completes the demonstration of \eqref{invers}:  $\Omega_{\omega^{-1}}=\Omega_{\omega}^{-1}$ and expression \eqref{Gammainv} is satisfied. \\

$\bullet$ {\sc \footnotesize  AN EQUIVALENCE RELATION DEFINED BY $\Omega_\omega$}\\

We say that the map $\Omega_\omega$ defined an equivalence relation between the space of operators defined on isomorphic Hilbert spaces.
For a given two hamiltonian operators $(H,H')$ we can define a relation between them 
\begin{align}\label{EQUIV-REL}
H'\sim H\Longleftrightarrow \exists \,\omega: H'=\Omega_{\omega}(H),
\end{align}
where $\Omega_{\omega}(H):=\omega H \omega^{-1}+i(\partial_t\,\omega)\omega^{-1}$ and $\omega$ is a nonsingular operator. The relation \eqref{EQUIV-REL} between two operators $H'\sim H$ is an equivalence in the sense that  for all operators $H,H',H''$ the following properties are true: 
\begin{align*}
(\pmb{R})\quad&H{\sim} H\quad (\mathit{reflexivity}) \\
(\pmb{S})\quad&H{\sim}H'\Longrightarrow H'{\sim} \,H\quad(\mathit{symmetry})\\
(\pmb{T})\quad&H''{\sim}\, H'\;{\scriptstyle\land} \; H'{\sim} \,H\Longrightarrow H''{\sim }\, H \quad (\mathit{transitivity})
\end{align*}
The first assertion ($\pmb{R}$) is true from the identity operator $\omega=I$ and by definition $\Omega_{I}(H)=H$. The assertion ($S$) is also true from the existence of the inverse operator $\omega^{-1}$ and construct through \eqref{invers} the \textit{inverse} connection  $H'\sim H$. The last assertion ($T$) is true from the composed transformation of non singular operator $\omega=\omega''\omega'$ and \eqref{composite}, such that $H''=\Omega_{\omega''}(H')$ and $H'=\Omega_{\omega'}(H)$, then $H''=\Omega_{\omega''}(\Omega_{\omega'}(H))%=\Omega_{\omega}''\omega'}(H)
=\Omega_{\omega}(H)$. Finally, we arrive to $H''\sim H$.

\vspace*{0.8cm}

\section*{\sc\pmb{A2. some required additional proofs}}

We start to prove \eqref{eq ham dep}, from $|\varphi(t)\rangle{=}\sum_{n\in I}\xi_n(t) |\varphi_n(t)\rangle$, taking the partial  time derivative and use the Schrödinger equation \eqref{schr eqt} $\sum_{n\in I} i(\partial_t\xi_n)\, |\varphi_n\rangle+i(\xi_n)\,|\partial_t\varphi_n\rangle{=}\sum_{n\in I}\lambda_n \xi_n\varphi_n$ then close from the left with a \textit{bra} $\langle\varphi_k|$ and using the orthonormal character of the dynamical basis $\pmb{\beta}_t$ finally obtain $ i\partial_t\xi_k +i\sum_{n{\in}I}\xi_n\langle\varphi_k|\partial_t\varphi_n\rangle{=}\lambda_k \xi_k$. Note that $\langle\varphi_m |\varphi_n\rangle{=}\delta_{mn}$ for all $t$, then $\langle\partial_t\varphi_m |\varphi_n\rangle{=-}\langle\varphi_m |\partial_t\varphi_n\rangle$. Finally we get $\langle\varphi_n |\partial_t\varphi_n\rangle{^*}{=}\langle\partial_t\varphi_n |\varphi_n\rangle{=-}\langle\varphi_n |\partial_t\varphi_n\rangle$, then $\int_0^t i\langle \varphi_n(s)|\partial_s\varphi_n(s)\rangle ds$ is a real quantity.

In order to arrive at \eqref{omega(H)} we start from 
\begin{align*}
\begin{aligned}
\Omega_\omega(H)&=\omega\,H\,\omega^\dagger + i(\partial_t\omega)\,\omega^{\dagger},\\
&=\mathlarger{\sum}_{n\in I}\pmb{\big{(}}\lambda_n+{i} (\partial_t\omega_n)\omega_n^*\pmb{\big{)}} \,|\psi_n\rangle \langle \psi_n|	\\
&+\;\mathlarger{\sum}_{(m,n)\in I^2}{i} \omega_m\omega_n^* \langle\partial_t\varphi_m|\varphi_n\rangle \,|\psi_m\rangle \langle \psi_n|,\\
\Omega_\omega(H)&=\mathlarger{\sum}_{n\in I}\pmb{\big{(}}\lambda_n - \partial_t f_n \pmb{\big{)}} \,|\psi_n\rangle \langle \psi_n|	\\
&+\;\mathlarger{\sum}_{(m,n)\in I^2}{i}\, e^{i(f_m-f_n)} \langle\partial_t\varphi_m|\varphi_n\rangle \,|\psi_m\rangle \langle \psi_n|,\\
\end{aligned}
\end{align*}
and extracting the term that corresponds to $n{=}m$  of the double  sum, finally obtain \eqref{omega(H)}.

To prove \eqref{additive hamiltonian} the procedure is as follows: using \eqref{idea primigenia} $\Omega_\omega(H{+}\widetilde{H}){=}\omega H\omega^{\dagger}{+}\Omega_\omega(\widetilde{H})$ and using the defined operator  ${\omega}$ then $\omega H\omega^{\dagger}{=}\sum_{n\in I}\lambda_n \,|\psi_n\rangle \langle \psi_n|$, so it only remains to find a hamiltonian $\widetilde{H}$ so that from the requirement established by \eqref{shortcut Omega} then $\Omega_\omega(\widetilde{H}){=}\sum_{n\in I} i\,\pmb{(}\langle \partial_t\varphi_n|\varphi_n\rangle\pmb{)}\,|\psi_n\rangle\langle\psi_n|$. Finally obtain this additive hamiltonian $\widetilde{H}$ via the inverse mapping of $(\Omega_\omega)^{-1}$ using \eqref{invers} we calculate
\begin{align*}
\begin{aligned}
\widetilde{H}&=\Omega_{\omega^{-1}}\pmb{\bigg{(}}\sum_{n\in I} i\,\pmb{(}\langle \partial_t\varphi_n|\varphi_n\rangle\pmb{)}\,|\psi_n\rangle\langle\psi_n|\pmb{\bigg{)}},\\
&=\sum_{n\in I} i\,\omega{^\dagger}\Big{[}\pmb{(} \langle \partial_t\varphi_n|\varphi_n\rangle\pmb{)}\,|\psi_n\rangle\langle\psi_n|\Big{]}\omega + i\partial_t (\omega^\dagger) \omega,\\
\widetilde{H}&=i\,\mathlarger{\sum}_{n\in I}\;|\partial_t\varphi_n\rangle \langle \varphi_n|+\pmb{\big{(}}\langle\partial_t\varphi_n|\varphi_n\rangle{-}i\partial_tf_n \pmb{\big{)}} \,{|\varphi_n\rangle} {\langle \varphi_n|},
\end{aligned}
\end{align*}
which is exactly  \eqref{additive hamiltonian}.

\end{document}